\documentclass[fleqn,10pt, twocolumn]{wlscirep}
\usepackage{setspace}
\usepackage{textcomp}
\title{Highly indistinguishable single photons from incoherently and coherently excited GaAs quantum dots} 

\author[1,*]{Marcus Reindl}
\author[2]{Jonas H. Weber}
\author[1]{Daniel Huber}
\author[1]{Christian Schimpf}
\author[1]{Saimon F. Covre da Silva}
\author[2]{Simone L. Portalupi}
\author[3]{Rinaldo Trotta}
\author[2]{Peter Michler}
\author[1,*]{Armando Rastelli}

\affil[1]{Institute of Semiconductor and Solid State Physics, Johannes Kepler University Linz, Altenbergerstraße 69, A-4040 Linz, Austria}
\affil[2]{Institut für Halbleiteroptik und Funktionelle Grenzflächen, Center for Integrated Quantum Science and Technology (IQ$^{ST}$) and SCoPE, University of Stuttgart, Allmandring 3, D-70569 Stuttgart, Germany}
\affil[3]{Dipartimento di Fisica, Sapienza Universita di Roma, Piazzale Aldo Moro 5, I-00185 Rome, Italy}

\affil[*]{Corresponding authors: marcus.reindl@jku.at, armando.rastelli@jku.at}

\begin{abstract}
Semiconductor quantum dots are converging towards the demanding requirements of photonic quantum technologies. Among different systems, quantum dots with dimensions exceeding the free-exciton Bohr radius are appealing because of their high oscillator strengths. While this property has received much attention in the context of cavity quantum electrodynamics, little is known about the degree of indistinguishability of single photons consecutively emitted by such dots and on the proper excitation schemes to achieve high indistinguishability. A prominent example is represented by GaAs quantum dots obtained by local droplet etching, which recently outperformed other systems as triggered sources of entangled photon pairs. On these dots, we compare different single-photon excitation mechanisms, and we find (i) a "phonon bottleneck" and poor indistinguishability for conventional excitation via excited states and (ii) photon indistinguishablilities above 90\% for both strictly resonant and for incoherent acoustic- and optical-phonon-assisted excitation. Among the excitation schemes, optical phonon-assisted excitation enables straightforward laser rejection without a compromise on the source brightness together with a high photon indistinguishability.
\end{abstract}
\begin{document}
\maketitle

\thispagestyle{empty}

\section{Introduction}
\label{sec:intro}
Quantum states of light are the unrivaled resource of future quantum communication networks. In the framework of single photon emission, epitaxial quantum dots (QDs), such as InGaAs QDs~\cite{somaschi.naturephot, PhysRevLett.116.020401} or InAsP QDs~\cite{reimer.naturecomm}, are well established solid-state sources. In such QDs, the exciton Bohr radius is typically larger than the spatial extensions of the semiconductor heterostructure and the wave function's physical properties are largely governed by the confinement potential. The opposite situation, the weak confinement regime, would mean that the Coloumb interaction between excitonic charge carriers begins to overwhelm effects related to the quantum confinement. This regime has attracted much attention in cavity quantum electrodynamics so as to reach the so-called strong-coupling regime~\cite{0268-1242-23-12-123001}. The reason is that the oscillator strength, i.e. light-matter coupling, in the weak confinement regime is largely enhanced~\cite{PhysRevB.60.13276}. Interestingly, the model system in these studies were already "natural" GaAs QDs formed via thickness fluctuations in thin GaAs/AlGaAs quantum wells~\cite{PhysRevLett.95.067401}. In terms of single photon emission properties, however, QDs in the weak confinement regime have received limited attention so far since these "natural" QDs provide poor control on the lateral confinement potential and feature a small energy separation between (discrete) confined states and (continuum) delocalized states. To address this issue while preserving large lateral extensions, different growth protocols have been developed over the years~\cite{PhysRevLett.92.166104, heyn.apl, huo.apl}. Here we focus on GaAs QDs obtained by droplet etching of nanoholes in AlGaAs followed by GaAs filling~\cite{huo.apl}, which typically have lateral sizes exceeding the free exciton Bohr radius in GaAs. The excitonic lifetime in this kind of dots ($\approx$250\,ps) is substantially shorter than the minimum lifetime expected for GaAs QDs in the strong confinement limit ($\approx$440\,ps)~\cite{PhysRevB.82.233302}, providing strong indication that excitons are weakly confined in our QDs. These QDs enabled the triggered emission of polarization entangled photons at near-unity fidelity~\cite{PhysRevLett.121.033902}, partly due to their large physical size which provides a dual advantage in this respect. First, the effect of residual anisotropies of the confinement potential on the excitonic fine-structure splitting (FSS) is strongly reduced compared to smaller QDs~\cite{PhysRevB.90.045430, PhysRevB.90.041304} and, second, the large decay rate alleviates effects of residual FSS and possible dephasing mechanisms leading to entanglement degradation~\cite{huber.naturecomm, keil.naturecomm}. Furthermore, under proper excitation conditions, GaAs QDs provided record low values of multi-photon emission probabilities~\cite{schweick.apl}, demonstrating that the weak confinement in these dots does not affect the single photon purity. Having these extraordinary properties at hand, all-photonic quantum teleportation~\cite{Reindl.scienceadv} and entanglement swapping~\cite{basso.swapp} schemes were already elaborated to possibly realize entanglement-based cryptography~\cite{PhysRevA.76.012307} using solid-state sources. Nonetheless, efficient long distance quantum communication~\cite{PhysRevLett.92.047904} demands highly indistinguishable photons and is also of paramount importance for photonic quantum information processing~\cite{Loredo:16,Vural:18}. In previous experiments on droplet-etched GaAs QDs based on strictly resonant two-photon excitation~\cite{huber.naturecomm, doi:10.1021/acs.nanolett.7b00777}, only limited values of photon indistinguishability ($V_{HOM}\approx 70\%$) have been usually observed. On the one hand, it is well known from InGaAs QD systems that near-optimal indistinguishable photons can be generated under strict resonant conditions~\cite{somaschi.naturephot,PhysRevLett.116.020401}. On the other hand, this usually requires to sacrifice source efficiency because of the delicate cross-polarized excitation/collection configuration needed  for laser rejection as well as freedom of choice regarding the photon polarization.

In this article we therefore investigate the limits of photon indistinguishability from droplet-etched GaAs QDs under various excitation conditions to assess the true emitter performance not yet revealed under two-photon pumping and, in particular, focus on practical, thus incoherent driving schemes. Thereby, we do not only compare strictly resonant excitation to the well known excitation via excited states~\cite{santori.nature} or the LA-phonon excitation~\cite{PhysRevB.90.241404} but also exploit the rarely considered triggered excitation via the LO-phonon~\cite{pooley.apl}, which yields excellent performance in the given material system. Our results demonstrate the possibility to generate highly indistinguishable photons even for a largely detuned excitation laser and allow insight into different relaxation processes present in droplet-etched GaAs QDs.

\section{Measurements and Results}
\label{sec:meandre}

\begin{figure*}[htbp]
\centering
\fbox{\includegraphics[width=\linewidth]{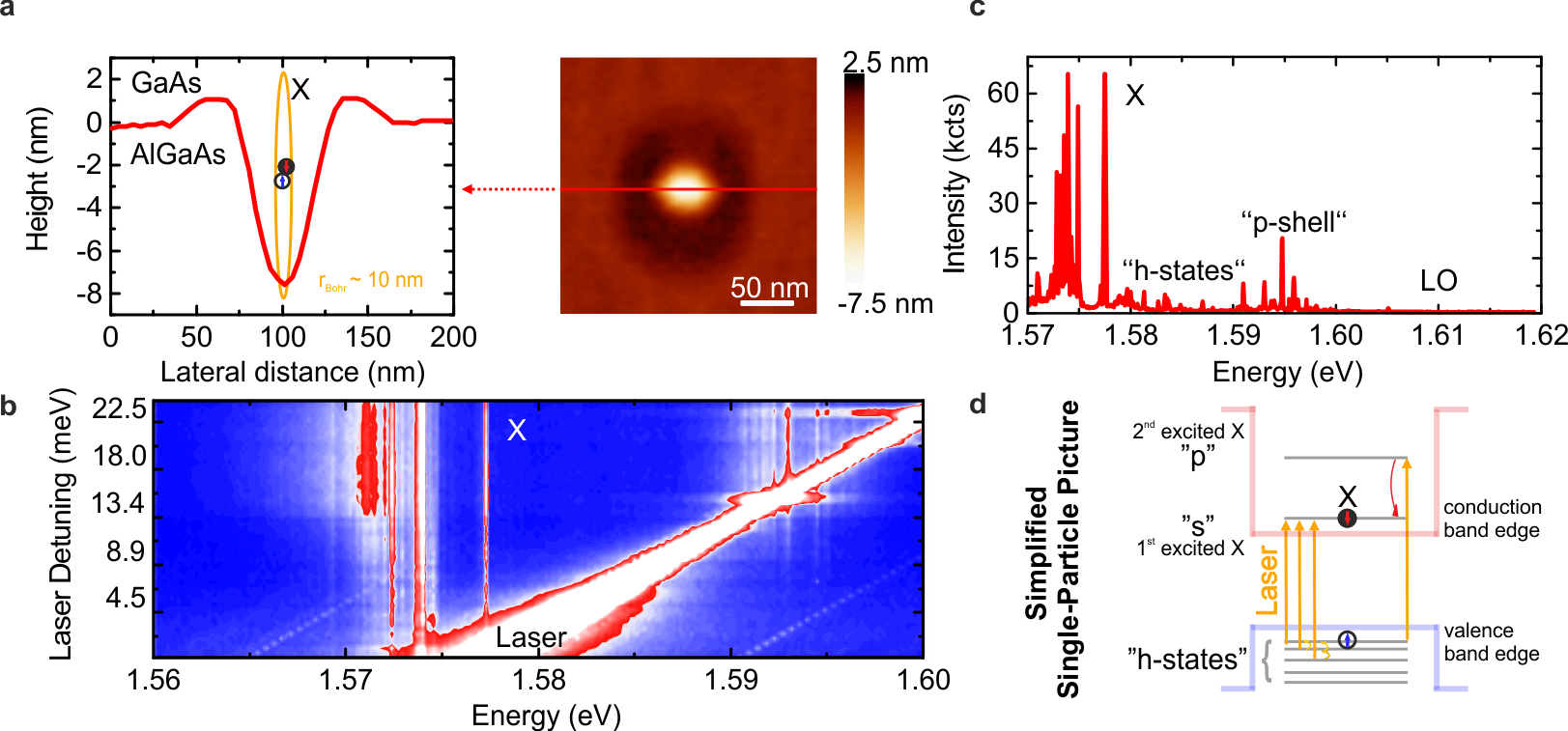}}
\caption{(a) Atomic force microscopy image of a droplet-etched nanohole in an AlGaAs layer. A line scan across the hole (red) is used to demonstrate the relative wavefunction extension of a created exciton with quoted Bohr radius $r_{Bohr}$. The actual GaAs quantum dot is obtained after filling the hole with GaAs and overgrowth with AlGaAs. (b) Photoluminescence excitation spectroscopy of a typical GaAs quantum dot for positive laser energy detunings with respect to the neutral exciton. (c) An above-band spectrum of a typical GaAs quantum dot under cw-excitation with a 488\,nm laser. (d) Single-particle sketch of the level structure to visualize the difference between the densely spaced hole states ("h-states") and the more confined conduction band s- and p-states.}
\label{fig:figure1}
\end{figure*}

We start our study via pulsed excitation with a laser energetically located exactly at the energy of the neutral exciton (X) of a typical droplet-etched GaAs QD, which is embedded in a low-Q DBR cavity hosting a solid immersion lens on top (see Supplemental Note 1 for more details). The relative extension of the created electron-hole pair - in terms of the Bohr radius - is depicted with the aid of atomic force microscopy (AFM) measurements in Figure \ref{fig:figure1}a to illustrate the weak, lateral confinement in our QDs. In general, all studies performed here are related to the neutral X. In a next step, we perform photoluminescence excitation spectroscopy (PLE) by detuning the energy of the excitation laser with respect to the X (see Figure \ref{fig:figure1}b) to possibly recognize any relevant resonant population mechanism of interest. By that, various effects are noticeable. First of all, under any positively detuned excitation condition, at least four other lines appear (for sufficient excitation power) on the low energy side of the X. We attribute the appearance of these lines to the excitation of the quantum dot in presence of extra carriers (most probably holes) stemming from residual doping. Because of the slow relaxation (see later in the text), this gives rise not only to ground-state trion emission but also to emission from trions with one extra carrier in an excited state. A further consequence of random charging is QD blinking, i.e. the suppression of resonant absorption (and hence X emission) when the QD is occupied by excess carriers~\cite{PhysRevB.97.195414}, an issue which can be solved by embedding the QD in diode structures or alleviated by weak above-bandgap illumination~\cite{doi:10.1021/acs.nanolett.7b00777, Gazzano:18}. Further detuning then reveals the presence of resonances (local maxima in the X intensity), while the number of low energy states is gradually increasing. A stable emission pattern is reached at a detuning energy of approximately 13\,meV, which we identify as the "p-shell" energy. In a single-particle picture, we would attribute the "p-shell" to a configuration featuring one electron in the first excited state in the conduction band. The appearance of several resonances between the electron s- and p-shell of the QD are instead attributed to densely spaced excited hole states~\cite{PhysRevLett.92.166104}. The presence of such excitonic states composed of ground-state electron and excited holes can be also seen from a spectrum taken in strong above-band excitation (see Figure \ref{fig:figure1}c).If the laser is exactly tuned to these "h-states", as schematically illustrated in Figure \ref{fig:figure1}d, we do observe maxima in the X intensity (see Supplementary Figure S1). We stress that the single-particle picture described above is a poor approximation for the large dots studied here and that resonances should be simply regarded as excited states of the system. In the opposite limit of weak confinement, we would interpret the densely spaced resonances as due to the center-of-mass motion of the exciton (dominated by the heavier mass of holes) in the lateral potential provided by the GaAs/AlGaAs QD and the "p-shell" to the internal excitation of the exciton (dominated by the lighter mass of the electron).  

\begin{figure}[htbp]
\centering
\fbox{\includegraphics[width=\linewidth]{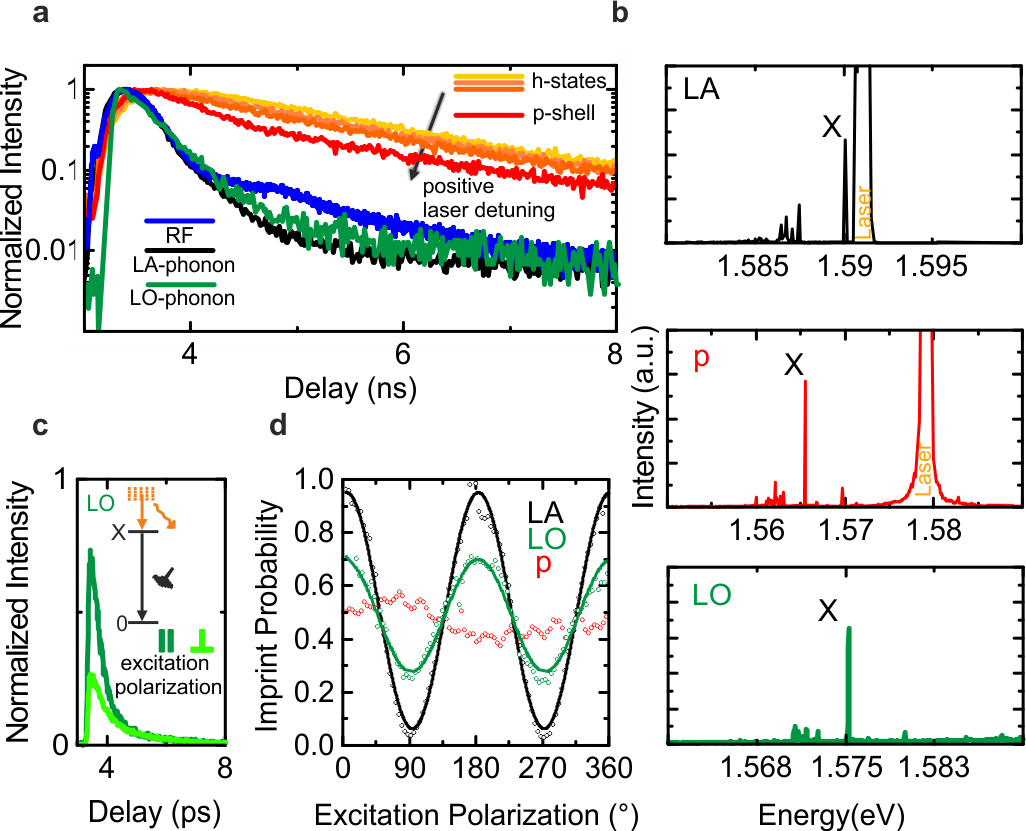}}
\caption{(a) Time-correlated single photon counting under various excitation schemes as measured on the X transition of a GaAs QD. The measurement under strict resonant excitation (blue) is performed on a different QD. The laser is detuned on the LA-phonon (black), the LO-phonon (green), on the "p-shell" (red) or the excited "h-states" (shades of orange). All decay times, except the data acquired under LO-phonon excitation, are extracted via a single exponential fit and under deconvolution of the instruments response function as well as the phase evolution of the fine structure split X states (only in RF). The data of the LO-phonon is instead fitted via a double exponential function. (b) Exemplary spectra of incoherently excited X states. The LA-phonon excitation (top, black), the p-shell excitation (center, red) and the LO-phonon excitation (bottom, green). (c) Decay time measurements on the LO-phonon for a parallel (dark green) and orthogonal (light green) configuration of excitation polarization with respect to the collection polarization, which is fixed to one of the two bright excitonic states. (Inset: Simplified sketch of the created polaron decaying via the emission of one LO-phonon and one X photon) (d) The probability of the polarization transfer from laser photons to the emitted single photons as a function of the excitation polarization angle. This angle is referred to a fixed polarization in the collection system aligned to one of the two bright excitonic states. The study is carried out for the resonant excitation conditions shown in (b), namely the LA-phonon (black), LO-phonon (green) and p-shell (red). The individual data points are fitted with a cosine-function.}
\label{fig:figure2}
\end{figure}

We continue our study with the evaluation of the recombination times of all detected X resonances. First of all, the "natural" resonant condition, the coherent excitation of the resonance fluorescence (RF)~\cite{doi:10.1063/1.5034402}, reveals a decay time of 209(5)\,ps under $\pi$-pulse condition (see Figure \ref{fig:figure2}a). The visible beating in the decay dynamics is associated to the phase evolution of the fine structure split bright excitons~\cite{Dada:16, eva.fluo}. In other words, we are not able to generate single photons with well-defined polarization due to the necessary polarization suppression. Therefore we investigate more practical incoherent population schemes for positively detuned excitation laser energies. A small detuning of 0.5\,meV allows us to properly dress the LA-phonon to directly populate the neutral X and yields a decay time of 236(8)\,ps (see Figure \ref{fig:figure2}a), a value compatible with that obtained under RF - considering that a different QD was used - or from previous studies under resonant two-photon excitation conditions~\cite{keil.naturecomm,huber.naturecomm}. An exemplary spectrum is shown in Figure \ref{fig:figure2}b and demonstrates a frequently encountered resonant behaviour of droplet-etched GaAs QDs: A strongly enhanced X transition accompanied by four suppressed, low energy lines. Surprisingly, further detuning towards the resonant condition of the first excited "h-state" on the same QD, just 5\,meV detuned from the X (see Supplementary Figure S1a), reveals a completely different behaviour. A much slower rise in the time trace followed by a decay time as long as 1.84(0.04)\,ns is observed. It is reasonable to assume that the actual radiative lifetime of the X transition did not change and that the decay is dominated by the slow relaxation of the excited state. This slow relaxation might be related to the fact that the energy separation between the excited- and ground-state ($\approx$ 5\,meV) is much smaller than the LO-phonon energy in GaAs (36\,meV~\cite{PhysRevB.83.041304}). Energy relaxation requires therefore a multitude of acoustic phonons, rendering the process much slower than typically observed for QDs in the strong confinement regime. For excited states with energy comparable to the LO-phonon energy the carrier relaxation times are enhanced due to an anharmonic polaron (carrier coupled to optical phonon) decay as reported in the literature~\cite{PhysRevB.76.241304, zibik.nature}. The thus observed "phonon bottleneck" is persistent even under resonant p-shell ($\approx$ 13\,meV) excitation (see Figure \ref{fig:figure2}b) in these QDs, where we witness decay times of 1.51(0.05)\,ns. This value is still a factor six slower than the time measured under s-shell and LA-phonon excitation. We note that the slow decay persists for optical and electrical above-bandgap excitation~\cite{PhysRevB.92.245439, doi:10.1021/acsphotonics.6b00935}. This makes us confident that the fast decay time ($\approx$200\,ps) measured under LA-phonon excitation coincides to the radiative decay time. The consistent slow decay time ($\approx$1.5-2\,ns) observed under excited state excitation sets in fact an upper limit to the contribution of possible non-radiative processes to the decay rate.

Since slow relaxation introduces a large time-jitter in the photon emission, we expect emission of photons with reasonable indistinguishability only under coherent s-shell and LA-phonon excitation. However, in these conditions the spectral proximity of the excitation laser is still not ideal. Parallel polarization of excitation and emission, as required for resonant excitation at maximal efficiency, is challenging~\cite{gerhardt.ellip, he.ellip}. In particular in terms of the single photon purity, which is a quantity highly vulnerable to scattered laser light. Thus, polarization filtering is usually applied at costs of the source efficiency (at least 50\% less) and arbitrary control of the polarization state is infeasible. A possible way to circumvent this problem is to detune the excitation laser even more, by approximately 36\,meV, the already mentioned LO-phonon energy in GaAs. Although the occurrence of resonant absorption at the LO-phonon energy is well known~\cite{PhysRevB.61.R10579, PhysRevLett.82.4114}, this pumping scheme has been rarely used for single photon emission~\cite{pooley.apl}, possibly because of the overlap of this resonance with the "p-shell" of smaller dots. A created polaron can now rapidly decay via the emission of one LO-phonon, leaving the QD populated with one exciton. This allows to populate the neutral X similar to the strictly resonant s-shell and LA-phonon-assisted excitation schemes, as proven by the measurement of the decay time in Figure \ref{fig:figure2}a, with the remarkable benefit of an excitation source being largely detuned with respect to the transition of interest. As a consequence, there is no need of any sophisticated filtering technique. An exemplary spectrum is shown in Figure \ref{fig:figure2}b, which resembles the resonant behaviour discussed beforehand. We want to emphasize at this point that the ideal resonant situation (strong X intensity compared to the above-mentioned low-energy lines) is usually not observed for all excitation conditions on a single QD at the same time. It means that an efficiently generated  X under LO-phonon excitation can show a non-ideal resonant behaviour under LA-phonon or p-shell excitation despite observing the same QD. In turn, this issue may depend on the details of the residual defects/doping configuration in the surroundings of the QD and may be solved once the QDs are embedded in diode structures. For this reason we show in Figure \ref{fig:figure2}b spectra of different QDs, which display clear resonances for each of the different excitation schemes. 

It is interesting to mention that any detuning from the LO-phonon resonance immediately leads to the reported stagnating recombination times and a similar emission pattern previously shown in Figure \ref{fig:figure1}b. From the decay curve obtained under LO-phonon excitation we extract a lifetime of 251(8)\,ps accompanied by a slow decay with characteristic time of 2.30(0.11)\,ns. Nevertheless, only a negligible amount of photons is found to be related to this slow decay channel (we do not account for photons lost to the low energy states emission), whose origin will be discussed later in the text. Most importantly, the incoherently placed laser now allows us to arbitrarily align the excitation polarization to one of the bright excitonic transition components separated by the FSS as demonstrated in Figure \ref{fig:figure2}c. Here, we first aligned the analyzing polarizer to one fine structure component of the X and then compared the temporal decay behaviour for parallel and orthogonal excitation configurations. The orthogonal case does not only reveal a weaker intensity but also a doubled decay time of 407(32)\,ps. In the ideal situation this anti-parallel driving scheme should be completely suppressed for the investigated X transition due to conservation of angular momentum, as the dissipated LO-phonon will carry no spin. However, the experimental result indicates the presence of depolarization mechanisms. For the sample used in the presented study, we can nevertheless "imprint" the laser polarization with a fidelity of $\sim$75\% (see Figure \ref{fig:figure2}d) on the desired transition component, meaning that we are at least 25\% more efficient than under strict resonant conditions, since no polarization rejecting elements are necessary. At least, because it is actually not trivial to drive a certain polarization axis under resonance fluorescence to reach the ideal 50\% efficiency condition. 

The LA-phonon assisted scheme exhibits a more efficient coupling to the two-level system (see Figure \ref{fig:figure2}d) as a consequence of the small energy difference between acoustic phonon bath and its dressed transition, which does not allow any relaxation via decay paths including excited states. Nonetheless, we want to emphasize that under optimal conditions (parallel polarization configuration) the scattering of laser light is unavoidable and prohibits perfect single photon properties of the emitter, a circumstance particularly related to the high pump energy needed in phonon-assisted excitation schemes.

\begin{figure}[htbp]
\centering
\fbox{\includegraphics[width=\linewidth]{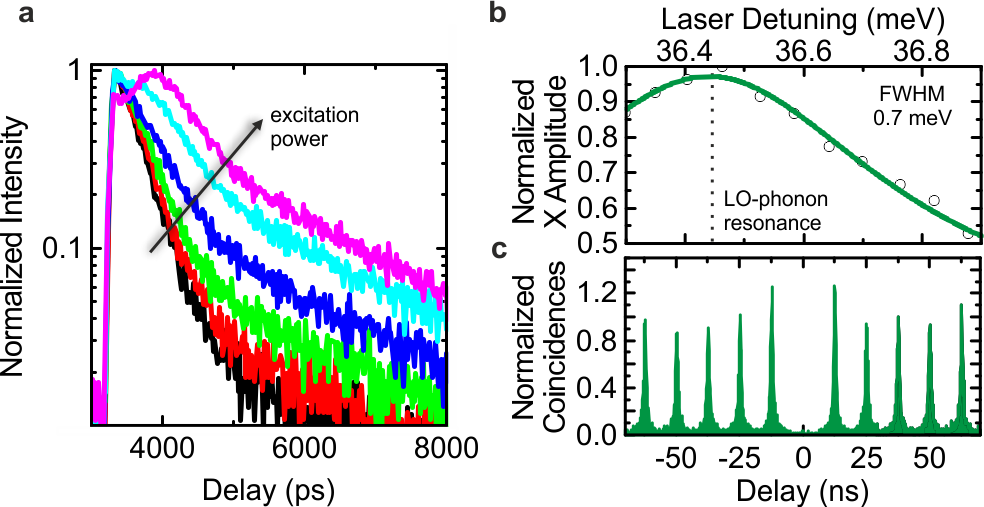}}
\caption{(a) Power dependent study of the X excitation and decay dynamics under (co-polarized) LO-phonon excitation. (b) Spectral width of the LO-phonon resonance as measured via detuning of the excitation laser. The data points are fitted by a Gaussian function to obtain the quoted FWHM. (c) Second-order correlation function measured on emitted X photons in a standard HBT setup under the LO-phonon excitation scheme.}
\label{fig:figure3}
\end{figure}

Before we compare the achievable photon indistinguishabilites of the presented excitation schemes, we want to investigate the LO-phonon excitation in our GaAs QDs in more detail. First of all, we study the power dependence of the decay dynamics represented in Figure \ref{fig:figure3}a. As we crank up the laser power, we can identify two distinct decay channels: (i) The desired decay of the created polaron via the ultrafast emission of one LO-phonon and the X photon (characterized by the fast rise and decay times) and, (ii), the undesired population of excited states present close to the LO-phonon energy, specifically within the spectral range of the used excitation laser, which then leads to a slow rise and similar recombination times previously seen under p-shell excitation. At excitation powers where the X intensity is identical to the ones observed under LA-phonon excitation or strict resonant condition, however, only a small fraction (<2\%) can be associated to this long decay channel. A more solid definition of the saturation intensity under LO-phonon excitation can not be given at this point as it would require a sophisticated model of the excitation power where the slow decay channel starts to affect the emission intensity and is beyond the scope of this work. 

Next, we tune the energy of the laser in small steps across the LO-phonon resonance to judge on its spectral width. Using a Gaussian fit (see Figure \ref{fig:figure3}b) we can estimate a FWHM of the LO-phonon resonance of 0.7(0.1)\,meV, which is of particular interest for a possible realization of the quantum interference involving remote QD sources~\cite{doi:10.1021/acs.nanolett.7b00777, weber.naturenano}. The LO-phonon energy itself is hereby found in a range of $\hbar\omega_{LO}=36.5\pm0.3$\,meV in our QDs. 

More importantly, in Figure \ref{fig:figure3}c we report on the Hanbury-Brown-Twiss (HBT) measurement on the X photons under the LO-phonon excitation and extract a multi-photon probability of $g^{(2)}_{X_{LO}}(0)=1.9(0.1)\times10^{-2}$, which is lower or equal than the multi-photon probabilities observed for the other excitation schemes under investigation (see Supplementary Figure S2). This value might be still deteriorated by the contribution of the long decay channel as observed in the second-order correlation measurement performed under p-shell excitation. Thus, one could expect the multi-photon emission probability to be further lowered once more ideal conditions for the exciton-phonon coupling are elaborated or by introducing moderate time gating during the measurement process. Another intriguing effect is the correlation statistics under the LO-phonon excitation on time scales comparable to the laser pulse frequency. At first glance, one might assume strong QD blinking to be present~\cite{PhysRevB.96.075430, PhysRevB.97.195414}, however, if we compare the second-order correlation statistics on long time scales to a truly blinking emitter possibly induced under p-shell excitation (see Supplementary Figure S3), we can conclude that this correlation effect origins from a different still unknown mechanism~\cite{Datta_2018}.

\begin{figure}[htbp]
\centering
\fbox{\includegraphics[width=\linewidth]{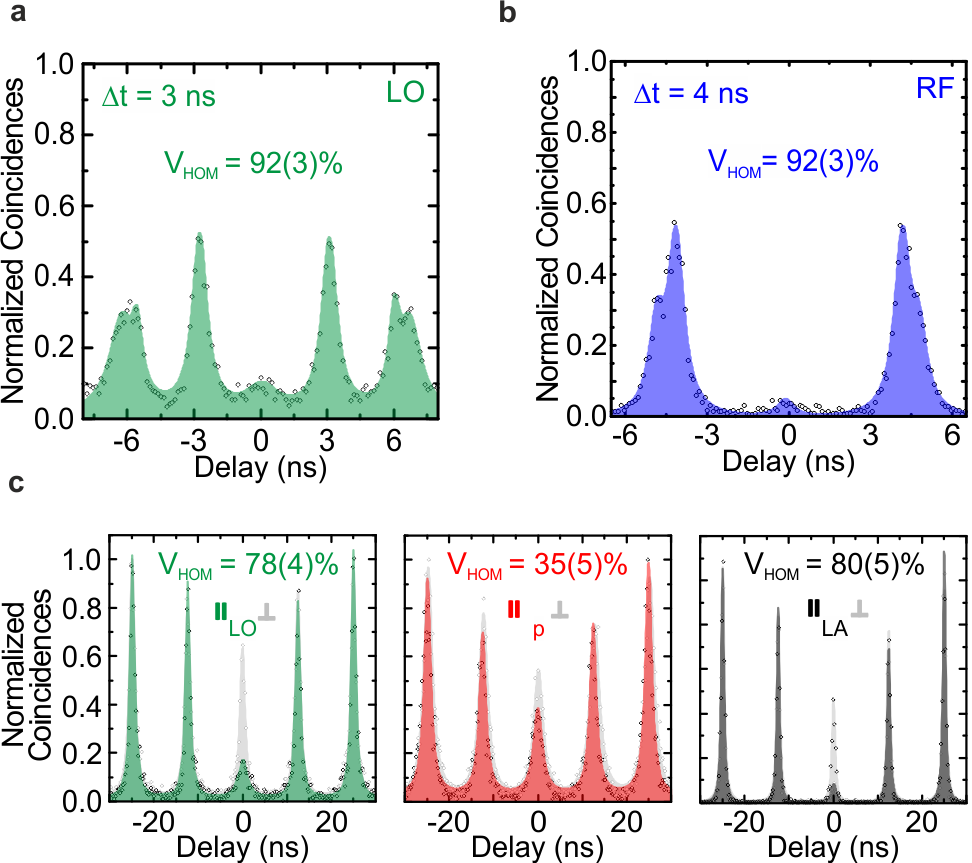}}
\caption{(a) Two-photon interference measurement on X photons using co-polarized settings under LO-phonon excitation. The time gap $\Delta$t between subsequently generated single photons used to probe the interference of the same source is 3\,ns. The quoted value of $V_{HOM}$ is obtained by fitting the resulting histograms with Lorentzian peaks (resulting in a necessary background substraction for the given $\Delta$t) and by taking into account the non-ideal properties of the interference beam-splitter. (b) Similar measurement but under strict resonant condition and slighlty increased $\Delta$t of 4\,ns between subsequently excited single photons. (c) Comparative study under LO-phonon excitation (green), p-shell excitation (red) and LA-phonon excitation (black) with $\Delta$t=12\,ns. All data points are again fitted assuming Lorentzian peaks to obtain the depicted values of the interference visibility. The cross-polarized setting (grey) is additionally depicted in each panel.}
\label{fig:figure4}
\end{figure}

Having these emission properties under LO-phonon excitation at hand, namely the close to zero multi-photon emission probability and fast transition time, we now present Hong-Ou-Mandel measurements among photons consequently emitted by the same QD. To this end we excite the QD with laser pulses separated by $\Delta$t=3\,ns and compensate this delay probabilistically with an unbalanced Mach-Zender interferometer. The result for the LO-phonon excitation is shown in Figure \ref{fig:figure4}a and reveals an interference visibility of $V^{HOM}_{X_{LO}}=92(3)\%$ compared to the case of completely distinguishable photons (next neighbouring peaks or cross-polarized measurements presented in Supplementary Figure S4). Additionally, the indistinguishability under strict resonant excitation (a corresponding spectrum is shown in Supplementary Figure S2) is reported in Figure \ref{fig:figure4}b to gauge the effect of external reservoir pumping, where a comparable visibility of $V^{HOM}_{X_{RF}}=92(3)\%$ (at larger excitation pulse separation of $\Delta$t=4\,ns) is observed. We want to emphasize that the evaluation of the visibility relies on a phenomenological fit of the histogram data by applying Lorentzian fits to the correlation peaks in both cases, as in~\cite{huber.naturecomm, doi:10.1021/acs.nanolett.7b00777}. A more physical approach would include a sophisticated model for the long decay channel in case of LO-phonon excitation and a proper evaluation of the FSS-induced state evolution present under resonance fluorescence. In view of the relatively large statistical uncertainties in the measurements, however, theoretical modelling (see~\cite{weber.naturenano}) returns visibility values comparable with Lorentzian fitting (see Supplementary Note 3 for details). The indistinguishability of X photons from several other QDs under LO-phonon exciation is reported in Supplementary Figure S5 to prove the general validity of our result. It shows that the dressing of LO-phonons induces a negligible amount of excitation time-jitter, since it can readily compete with values observed under strict resonant condition and is on par to the best results reported for conventional InGaAs QDs~\cite{he.naturenano} under resonance fluorescence. This result is important bearing in mind that the value is achieved without Purcell enhancement or phonon sideband filtering~\cite{smith.naturephot}. Cavity structures featuring Purcell enhancement and phonon sideband filtering could therefore enable near-unity indistinguishable photon emission even for the LO-phonon excitation scheme. Furthermore, the indistinguishability values are significantly higher than those reported for InGaAs QDs under LO-phonon excitation ($\approx$70\%~\cite{pooley.apl}). We tentatively attribute this circumstance to the high transition rates of our weakly confining QDs, which alleviate dephasing effects originated from phonon interaction and/or spin noise. However, we may expect similar performance also for strongly confined QD systems if embedded in Purcell-enhanced cavities, meaning that the practical excitation scheme studied here is generally not restricted to weakly confined QDs.

In a last step, we want to compare the LO-phonon excitation method to the other possible incoherent population schemes in our GaAs QDs. The time gap between subsequently excited single photons within the HOM interferometer is thereby enhanced to 12\,ns, which allows us to properly compare the visibility also in the case of a slow decay, i.e. observed under p-shell excitation. First, we repeat the measurement on the LO-phonon (see Figure \ref{fig:figure4}c) and extract a visibility of $V^{HOM}_{X_{LO}}=78(4)\%$. We do observe a degradation of the indistinguishability, which we attribute to charge noise, as known from the literature to be often present also on these timescales~\cite{huber.naturecomm,PhysRevLett.116.033601}. It is particularly present in structures which do not alleviate the large bandwidth noise attributed to excess charge carriers in the QD environment via Purcell enhancement~\cite{PhysRevLett.116.020401}. A different picture is revealed in Figure \ref{fig:figure4}c under p-shell excitation. While under certain detuning conditions the spectra are dominated by the X transition and strongly resemble those obtained under LO-phonon excitation, the very long relaxation time and induced time-jitter leads to photon indistinguishability values barely exceeding 30\%. This pronounced degradation has not been observed in InGaAs systems, where even under p-shell excitation the emission of indistinguishable photons ($V_{HOM}\approx 90\%$~\cite{PhysRevB.96.165306}) is realizable because of a slower radiative recombination rate combined with faster relaxation of the excited states. This allows us to conclude that, different from QDs featuring strong confinement, "p-shell" excitation is not a viable route to obtain photons with meaningful indistinguishability from weakly confining QDs. Instead, we focus once more on the resonant LA-phonon excitation. In Figure \ref{fig:figure4}c we report on a visibility of $V^{HOM}_{X_{LA}}=80(5)\%$, compatible to the value obtained under LO-phonon excitation, which demonstrates that it is practically irrelevant which phonon decay channels are involved in droplet-etched GaAs QDs to obtain triggered indistinguishable single photons and underlines the advantage of the LO-phonon-assisted excitation to be considered in quantum network applications.

\section{Conclusion}
\label{sec:conc}

We have evaluated possible incoherent excitation mechanisms of the X state in droplet-etched GaAs QDs to determine suitable excitation schemes to obtain the emission of single and indistinguishable photons and also compare them to the coherent resonance fluorescence. While we found limited performance in excitation schemes relying on the population of excited states, which we explain by the fact that the excitons in our flat QDs are weakly confined and efficient relaxation through LO-phonon emission is hindered, resonant phonon-assisted excitation schemes are instead capable to deliver highly indistinguishable single photons. We expect our findings to be relevant also to other QD systems and attribute the excellent performance under phonon-assisted excitation to the enhanced oscillator strength present in our weakly confined QD system~\cite{PhysRevA.69.032305}. The largely detuned (with respect to the neutral X) LO-phonon energy can act as a coupling interface to generate indistinguishable single photons in a robust way. Furthermore, a more precise dressing of the LO-phonon may eventually suppress the long decay channel associated to the undesired population of excited states. This might be realizable once the exciton-phonon coupling is further engineered, which may be possible via optimization of the QD size, or likewise, the exciton wavefunction~\cite{PhysRevLett.83.4654}. This may be feasible either by engineering the QD structural properties or by introducing elastic stress, as already done previously on droplet-etched GaAs QDs~\cite{kumar.apl,doi:10.1021/acsphotonics.6b00935,xueyong.naturecomm}. The establishment of QD cavity systems~\cite{PhysRevB.98.045309} and devices which allow the reduction of charge noise~\cite{PhysRevLett.101.226603} to ensure indistinguishable photons on arbitrary time scales~\cite{PhysRevLett.116.213601} can then position GaAs QDs as excellent sources of upcoming quantum photonic networks offering practicable excitation schemes..

\section*{Funding Information}

This work was financially supported by the Austrian Science Fund (FWF): P 29603, the Linz Institute of Technology, the European Research Council (ERC) under the European Unions Horizon 2020 research and innovation programme (SPQRel, Grant agreement No. 679183) and European Union Seventh Framework Programme (FP7/2007-2013) under grant agreement no. 601126 (HANAS). The authors would like to thank the DFG for financial support via the project Mi500/27-1. The research
of the IQ$^{ST}$ was financially supported by the Ministry of Science, Research and Arts Baden-Württemberg.

\section*{Acknowledgments}

We acknowledge fruitful discussions with Y. H. Huo, M. Munsch, R. Warburton, G. Weihs, K. D. Jöns, and P. Klenovský.\\

See Supplement 1 for supporting content.

\bibliography{references}

\end{document}